\documentclass[10pt,conference]{IEEEtran}
\usepackage{amsmath,amsfonts}
\usepackage{amssymb}
\usepackage{algorithmic}
\usepackage{algorithm}
\usepackage{array}
\usepackage{makecell} 
\usepackage[top=0.70in,bottom=1.00in,left=0.65in,right=0.65in,columnsep=0.27in]{geometry}  
\usepackage[caption = false]{subfig}
\usepackage{textcomp}
\usepackage{stfloats}

\usepackage{url}
\usepackage{verbatim}
\usepackage{graphicx}
\usepackage{cite}
\usepackage{bm}
\usepackage{amsmath}  
\usepackage{mathrsfs}  
\usepackage{xcolor}

\begin{document}

\title{Graph Neural Network-based End-to-End Learning for Multi-User MIMO Systems}

\author{\IEEEauthorblockN{Hao Chang\IEEEauthorrefmark{1}, Hoang Triet Vo\IEEEauthorrefmark{1}, 
Alva Kosasih\IEEEauthorrefmark{2}, Branka Vucetic\IEEEauthorrefmark{1}, Wibowo Hardjawana\IEEEauthorrefmark{1}}\\
\IEEEauthorblockA{\IEEEauthorrefmark{1}School of Electrical and Computer Engineering, The University of Sydney, Sydney, Australia.\\
\IEEEauthorrefmark{2}Nokia Technology Standards, Finland\\
\{hao.chang, branka.vucetic, wibowo.hardjawana\}@sydney.edu.au, hovo0651@uni.sydney.edu.au, and alva.kosasih@nokia.com}}

\maketitle

\begin{abstract}   
End-to-end (E2E) learning has recently been proposed to jointly design the modulator and symbol detector by using deep neural networks (DNNs). However, existing schemes lack sufficient capability to cancel multi-user interference (MUI) in uplink multi-user multiple-input multiple-output (MU-MIMO) systems.
In this paper, we propose a graph neural network (GNN)-based E2E learning scheme that employs a GNN-based modulator to generate learned constellation points, and a GNN-based detector to cancel MUI. They are jointly optimized to minimize the symbol error rate (SER) performance loss.
Simulation results demonstrate that the proposed E2E outperforms existing schemes with a predefined modulator. Specifically, it achieves an approximate 2 dB gain in a high MUI environment and surpasses even the maximum-likelihood (ML) detector in a low MUI condition.

\end{abstract}


\section{Introduction}
\label{S_Intro}

Multi-user multiple-input multiple-output (MU-MIMO) systems have attracted much attention as an essential technique for 5G and beyond networks to increase spectrum efficiency. In the uplink MU-MIMO system, multiple single-antenna users transmit data simultaneously to a base station with multiple antennas via the same wireless spectrum, leading to high spectral efficiency. However, this simultaneous transmission causes multi-user interference (MUI), significantly degrading the system's symbol error rate (SER) performance.   

Recently, a specific type of deep neural networks (DNN), graph neural network (GNN), has attracted enormous attention in communication networks \cite{9846958}, particularly for symbol detector design for MU-MIMO systems \cite{Alva_GEP}.  
The MUI relationship is encoded as edges between nodes that represent users in the GNN graph. Messages computed by using DNN at each node and edge are propagated and aggregated on the graph, with the final messages used to predict the symbols. The MUI relationship encoding in the graph is what differentiates GNN from standard DNN approaches.
To date, a combination of classical state-of-the-art iterative expectation propagation (EP) \cite{Jespedes-TCOM14} and GNN \cite{AScotti_GNN_2020} MIMO symbol detectors has shown the best SER performances\cite{Alva_GEP,10306259}. In the above schemes, the GNN is used to refine the cavity probability estimation of symbols, which subsequently improves the posterior estimation in each EP iteration.
These detection schemes use a predefined modulator at the transmitter. The modulator maps the user messages to complex symbols, referred to as constellation points. The constellation points are arranged in a predefined structure. These symbol detection schemes with a predefined modulator in \cite{Alva_GEP,10306259} yield suboptimal performance compared to end-to-end (E2E) learning, where the structure of constellation points in the modulator at the transmitter and the detector at the receiver are jointly designed.  
E2E learning has recently been proposed in \cite{Timothy_intro,9735143} using a DNN-based modulator to map the user messages to transmitted complex symbols at the transmitter, and a DNN-based detector to map the received symbols to the estimated user messages at the receiver. The parameters in these two DNNs are jointly optimized by minimizing the E2E loss function.  
The above schemes only consider multi-user or MIMO systems with a small number of users and antennas, and the use of multi-layer perceptrons (MLPs) in their detectors lacks sufficient modeling capability to capture the MUI, as the MLP does not model the interference relationships. 
To date, no work has been conducted on E2E learning that can capture complex MUI relationships in MU-MIMO systems.
 
In this paper, we propose a GNN-based E2E learning scheme for uplink MU-MIMO systems, consisting of a GNN-based modulator ($\rm GNN^{tx}$) and a GNN-based detector ($\rm GNN^{rx}$).  
The reason we use GNN at both the transmitter and receiver is that the GNN has the ability to model the relationships between nodes via edges in the graph. In $\rm GNN^{tx}$, nodes represent constellation points and edges capture the Euclidean distance between them, whereas in $\rm GNN^{rx}$, nodes represent user symbols and edges capture the MUI.
During offline training, initial constellation points are input to $\rm GNN^{tx}$ for the initial node and edge features. After the iterative message passing in $\rm GNN^{tx}$, the final node feature after normalization is the learned constellation points, which are used for the mapping between the user message and transmitted complex symbol. The $\rm GNN^{rx}$ adapted from the GEPNet \cite{Alva_GEP} is used to estimate the user message based on the learned constellation, received signal, and channel state information (CSI). $\rm GNN^{rx}$ consists of EP and GNN modules, and they are executed iteratively. During each EP iteration, GNN takes the EP's cavity probability estimation of user symbols as input and outputs a refined cavity probability of the user message via iterative message passing. Unlike GEPNet \cite{Alva_GEP}, in which GNN estimates real and imaginary parts of complex symbols (user messages), our GNN design directly outputs complex symbol probability. 
The refined cavity probability of the user message is then used by EP for calculating the posterior probability estimation of user symbols, completing one EP iteration.  
The $\rm GNN^{rx}$'s iterative process repeats, and the final estimated cavity probability of the user message from GNN is used to calculate the cross-entropy (CE) loss.
All NN parameters in both $\rm GNN^{tx}$ and $\rm GNN^{rx}$ are jointly optimized to minimize the CE loss between the estimated and true user messages via backpropagation. 
After the offline training is completed, the final learned constellation from $\rm GNN^{tx}$ is then broadcast to all users by the base station.
The E2E system is now ready for online inference, with users transmitting messages using the final learned constellation and the base station employing $\rm GNN^{rx}$ with optimized NN parameters and constellation knowledge to estimate user messages (complex symbols). 
Simulation results demonstrate that the proposed E2E outperforms existing schemes with a predefined modulator. Specifically, it outperforms \cite{Alva_GEP} by approximately 2 dB at SER $=10^{-4}$ in a high MUI environment and even surpasses the maximum-likelihood (ML) in a low MUI condition. This performance improvement is achieved with a complexity comparable to that of \cite{Alva_GEP} during online inference.


Our main contribution is the development of a GNN-based E2E learning framework for MU-MIMO systems that jointly optimizes the modulator and detector, resulting in better SER performance than other schemes, as demonstrated by the simulation results. The proposed E2E addresses the limitations in \cite{Alva_GEP}, which does not optimize the modulator together with the detector, and in \cite{Timothy_intro}, which lacks sufficient modeling capability to capture the MUI.

{\bf Notations}: $a$, $\mathbf{a}$, and $\mathbf{A}$ denote scalar, vector, and matrix, respectively.
$\mathbf{A}^{\top}$ represents the transpose of matrix $\mathbf{A}$. 
$\mathbb{C}^{M\times N}$ and $\mathbb{R}^{M\times N}$ denote an $M\times N$ complex-valued and real-valued matrix, respectively. $\mathbf{I}_M$ represents an $M\times M$ identity matrix. 
$\|\mathbf{a}\|$ denotes the Frobenius norm of $\mathbf{a}$. 
We use $\mathcal{N}(\mathbf{x}:\boldsymbol{\mu},\boldsymbol{\Sigma})$ to represent a multi-variate Gaussian distribution for a random variable $\mathbf{x}$ with mean $\boldsymbol{\mu}$ and covariance $\boldsymbol{\Sigma}$.  
$\Re(\cdot)$ and $\Im(\cdot)$ are operators to take the real and imaginary parts, respectively. 
We use $\left[a,b\right]$ to represent the concatenation of $a$ and $b$. We use $a^{\rm tx}$ and $a^{\rm rx}$ to represent the variable $a$ at the transmitter and receiver sides, respectively.

\section{GNN preliminaries}
\label{S_gepnet}      
In this section, we briefly introduce the message passing-based GNN proposed in \cite{9048920}, which is then used for symbol detection in \cite{AScotti_GNN_2020,Alva_GEP}, consisting of propagation, aggregation, and readout modules. This GNN architecture is then used for the proposed E2E in Section IV. For a given graph with $N$ nodes, the feature of node $i$ at GNN iteration $l\in\{1,\dots,L\}$ is defined as $\mathbf{u}_i^{(l)}\in \mathbb{R}^{S_u}$, and the edge feature that represents the relationship from node $i$ to $j$ is $\mathbf{e}_{ij}$.
The value of each entry in these two vectors represents a feature used by nodes and edges in GNN, respectively.

\paragraph{Propagation}
In the propagation module, each node propagates a message to its neighboring nodes.
For each pair of nodes $i$ and $j$, the message passed from node $i$ to node $j$ at iteration $l$ captures the interference from node $i$ to $j$, expressed as
 \begin{equation}\label{eq_prop}
 \mathbf{m}_{i\rightarrow j}^{(l)} =\mathcal{P} \left( \mathbf{u}_j^{(l-1)}, \mathbf{u}_i^{(l-1)}, \mathbf{e}_{ij} \right),
 \end{equation}
where $\mathcal{P}$ is an MLP with three layers, and the hidden layer sizes are $N_{p_1}, N_{p_2}$, and the output layer size is $S_u$. 
Rectified linear unit (ReLU) is used after each hidden layer. $\mathbf{u}_i^{(l-1)}$ and $\mathbf{u}_j^{(l-1)}$ represent the feature of node $i$ and $j$ at GNN iteration $l-1$, respectively.
\paragraph{Aggregation}
Each node's feature is updated based on the incoming messages, a gated recurrent unit (GRU) network \cite{cho2014learning}, and a single-layer NN, given as
\begin{subequations}\label{eq_agg}
            \begin{equation}\label{eq_agg1}
 \mathbf{g}_j^{(l)} = \mathcal{U}  \left( \mathbf{g}_j^{(l-1)},\sum_{{i=1,i \neq j}}^{N} \mathbf{m}_{i\rightarrow j}^{(l)}, \mathbf{r}_j  \right),
            \end{equation}
            \begin{equation} \label{eq_agg2}
\mathbf{u}_j^{(l)}= \mathbf{W} \cdot \mathbf{g}_j^{(l)} + \mathbf{b},
            \end{equation}
\end{subequations}
where the function $\mathcal{U}$ represents the GRU, whose current and previous hidden states are $\mathbf{g}_j^{(l)}, \mathbf{g}_j^{(l-1)} \in \mathbb{R}^{N_{u_1} }$, respectively. The reason we use GRU is to capture the information between two consecutive iterations, as the messages are correlated between these iterations.
$\mathbf{r}_j$ represents the prior information for node $j$, which is adopted in \cite{Alva_GEP}.
After the GRU, a single-layer NN with a weight matrix $\mathbf{W} \in \mathbb{R}^{S_u \times N_{u_1}}$, and bias vector $\mathbf{b} \in \mathbb{R}^{S_u }$ are used to generate the updated node feature.  
The propagation and aggregation processes are repeated $L$ times to produce the final node feature $\mathbf{u}_j^{(L)} \in \mathbb{R}^{S_u}$.
\paragraph{Readout}
The readout module is used to generate the estimated probabilities of nodes based on the final node features, expressed as
\begin{subequations}\label{eq_readout}
    \begin{equation}
    \mathbf{\bar{p}}_{j} = \mathcal{R}(\mathbf{u}_{j}^{(L)}),
    \end{equation} 
    \begin{equation}
    p_{j,m} = \frac{{\sf exp} (\bar{p}_{j,m}) }{ \sum_{m=1}^{M}{\sf exp} (\bar{p}_{j,m} )},
    \end{equation} 
\end{subequations}
where $\mathbf{\bar{p}}_{j}\!\!=\!\![\bar{p}_{j,1},\dots,\bar{p}_{j,m},\dots,\bar{p}_{j,M}]\!\in\!\mathbb{R}^{M}$ and $\mathbf{p}_{j}\!\!=\!\![p_{j,1},\dots,p_{j,m},\dots,p_{j,M}]\!\in\!\mathbb{R}^{M}$ represent the unnormalized and normalized symbol probabilities for all $M$ states for node $j$, respectively. 
$\mathcal{R}$ is a three-layer MLP with sizes $N_{r_1}, N_{r_2}$ for hidden layers and $M$ for the output layer, respectively, and ReLU is used after each hidden layer.

\section{System Model}
\label{S_sys_mod} 
    
We consider an uplink MU-MIMO system in which $N_t$ users, each with a single antenna, communicate to a base station equipped with $N_r$ antennas.
At the transmitter side, each user generates $\log_2{M}$ information bits, which are represented by message $s\in \{1,\dots,M\}$. We define $\mathbf{s}=[s_1,\dots,s_k,\dots,s_{N_t}]^{\top}$, where $s_k, k\in \{1,\dots,N_t\}$ is the message from user $k$.  
The modulator is used to map the user message $s_k$ to a complex symbol $\tilde{x}_k$ by using one of the possible $M$ constellation points $a_{s_k}$, which is chosen from the complex-valued constellation set $\boldsymbol{\Omega} = [a_1, \dots,a_m,\dots, a_M] ^{\top}\in \mathbb{C}^{M}$ under the power constraint of $(\sum^M_{m=1}|a_m|^2)/M = 1$. 
The modulated symbols are then transmitted over the wireless channel. The corresponding received signal is then given by
\begin{equation} \label{eq_yhx_complex}
\tilde{\mathbf{y}} = \tilde{\mathbf{H}} \tilde{\mathbf{x}} + \tilde{\mathbf{n}},
\end{equation}
where $\tilde{\mathbf{y}}=[\tilde{y}_1, \ldots, \tilde{y}_{N_r}]^{\top} \in \mathbb{C}^{N_r} $ is the received signal, $\tilde{\mathbf{x}} = [\tilde{x}_1, \cdots, \tilde{x}_{N_t}]^{\top}\in \mathbb{C}^{N_t}$ is the transmitted signal that consists of modulated complex symbols. 
$\tilde{\mathbf{n}} \sim \mathcal{N}( 0,\tilde{\sigma}^2\mathbf{I}_{N_r})\in \mathbb{C}^{N_r}$ is the additive white Gaussian noise (AWGN), and $\tilde{\sigma}^2$ is the noise variance. $\tilde{\mathbf{H}}=[\tilde{\mathbf{h}}_1,\dots, \tilde{\mathbf{h}}_k, \ldots, \tilde{\mathbf{h}}_{N_t}]  \in \mathbb{C}^{N_r \times N_t} $ is the Rayleigh fading channel matrix between $N_r$ receive antennas and $N_t$ users. $\tilde{\mathbf{h}}_k$ is the $k$-th column vector of $\tilde{\mathbf{H}}$ that denotes wireless channel coefficients between the received antennas and the $k$-th user, and $\tilde{\mathbf{h}}_k \sim \mathcal{N}(0,\frac{1}{N_r}\mathbf{I}_{N_r})$.

We define a real-equivalent model of \eqref{eq_yhx_complex} such that
$\mathbf{x} = [\Re(\tilde{\mathbf{x}})^{\top}  \Im(\tilde{\mathbf{x}})^{\top}]^{\top} =[x_1,\dots,x_K\dots,x_{2N_t}]^{\top} \in \mathbb{R}^{2N_t}$ 
$\mathbf{y} = [\Re(\tilde{\mathbf{y}})^{\top}  \Im(\tilde{\mathbf{y}})^{\top}]^{\top} \in \mathbb{R}^{2N_r}$, $\mathbf{n} = [\Re(\tilde{\mathbf{n}})^{\top} \Im(\tilde{\mathbf{n}})^{\top}]^{\top} \in \mathbb{R}^{2N_r}$, and 
$\mathbf{H}=\begin{bmatrix}
  \Re(\tilde{\mathbf{H}})\!\!\!\!\! & -\Im(\tilde{\mathbf{H}}) \\ 
  \Im(\tilde{\mathbf{H}})\!\!\!\!\!  & \Re(\tilde{\mathbf{H}}) 
\end{bmatrix} \in \mathbb{R}^{{2N_r} \times {2N_t}}$.
The real-equivalent model of \eqref{eq_yhx_complex} can then be written as
\begin{equation} \label{eq_yhx_real}
\mathbf{y} = \mathbf{H} \mathbf{x} + \mathbf{n},
\end{equation}
where the covariance of $\mathbf{n}$ is $\sigma^2\mathbf{I}_{2N_r}, \sigma^2=\tilde{\sigma}^2/2$. 
We define the signal-to-noise ratio (SNR) $ = 10\log \frac{\mathbb{E} \left[ \|\mathbf{H}\mathbf{x} \|^2 \right]}{\mathbb{E} \left[ \|\mathbf{n} \|^2 \right]}$dB. 
$\mathbf{H}$ and $\sigma^2$ are assumed to be perfectly known at the receiver side. We use the system model \eqref{eq_yhx_real} for the E2E system.

\section{E2E architecture for offline training}
\label{s_e2e_train}
\begin{figure*}
\centering
{\includegraphics[width=\linewidth]{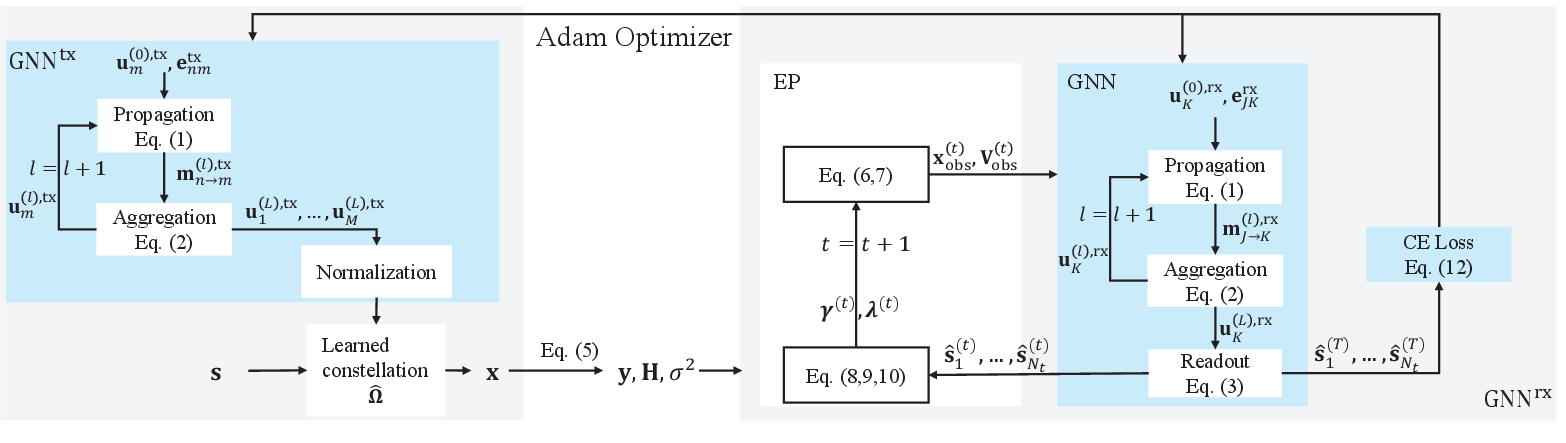}}   
\vspace{-0.8cm}
\caption{Proposed E2E architecture for offline training}   
\vspace{-0.5cm}
\label{fig_e2e_train}
\end{figure*} 

We train the E2E system offline. In 5G NR systems, this process can be performed in the cloud or a data center. In the following subsections, we present the proposed E2E architecture for offline training at the transmitter and receiver sides, using GNN architecture introduced in Section II. The proposed E2E architecture is shown in Fig. \ref{fig_e2e_train}.

\subsection{Learning at the Transmitter Side}
\label{learn_transmitter}

At the transmitter side, we propose a GNN-based modulator, i.e., $\rm{GNN}^{tx}$, to generate learned $M$ constellation points, which are used for the mapping between user messages and transmitted symbols, as shown in Fig. \ref{fig_e2e_train}.
The $\rm{GNN}^{tx}$ follows the GNN architecture in Section II, where each node $m\in\{1,\dots,M\}$ represents one constellation point and the edge $\mathbf{e}_{nm}^{\rm{tx}}, n\neq m$ represent the Euclidean distance between node $n$ and $m$. We define the node feature for the $m$-th constellation point at iteration $l\in\{1,\dots,L\}$ as $\mathbf{u}_{m}^{(l),\rm{tx}}$, and the edge feature from the $n$-th to the $m$-th points as $\mathbf{e}_{nm}^{\rm{tx}}$. These features are initialized as $\mathbf{u}_{m}^{(0),\rm{tx}}=[\Re(a_m),\Im(a_m)]^{\top}\in \mathbb{R}^{2}$ and $\mathbf{e}_{nm}^{\rm{tx}}=[|a_n-a_m|/\max(|a_n-a_m|),\sigma^2]^{\top}$, where $n,m\in\{1,\dots,M\}$.
The $\rm{GNN}^{tx}$ then repeats the propagation \eqref{eq_prop} and aggregation \eqref{eq_agg} processes $L$ times. The prior information $\mathbf{r}_m$ in \eqref{eq_agg} is not used.
The final node features $\mathbf{u}_{1}^{(L),\rm{tx}},\dots,\mathbf{u}_{M}^{(L),\rm{tx}}$ are the unnormalized $M$ constellation points, and $S_u^{\rm{tx}}=2$ is the size of node feature. Here, we use $S_u^{\rm{tx}}$ instead of $S_u$ to indicate that it's a variable at the transmitter. 
We then use a normalization module to ensure the learned constellation points $\hat{a}_m$ have a unit average power by applying $\hat{a}_m=\mathbf{u}_{m}^{(L),\rm{tx}}[0]/P+j\mathbf{u}_{m}^{(L),\rm{tx}}[1]/P$, and $P=\sqrt{\sum^M_{m=1}\|\mathbf{u}_{m}^{(L),\rm{tx}}\|^2/M}$. $\mathbf{u}_{m}^{(L),\rm{tx}}[0]$ and $\mathbf{u}_{m}^{(L),\rm{tx}}[1]$ represent the feature of the real and imaginary part of the learned constellation $\hat{a}_m$. Finally, the learned constellation points are $\hat{\boldsymbol{\Omega}}=[\hat{a}_{1},\dots,\hat{a}_{M}]\in \mathbb{C}^{M}$. 

After that, we map the message $s_k$ of user $k$ to the modulated symbol via the learned constellation $\hat{\boldsymbol{\Omega}}$. The transmitted modulated symbol is the $s_k$-th entry of $\hat{\boldsymbol{\Omega}}$, denoted as $\tilde{x}_k=\hat{a}_{s_k}$. The modulated symbols of all users are modeled as $\mathbf{x}$ in \eqref{eq_yhx_real} and transmitted via the real-equivalent channel $\mathbf{H}$.

\subsection{Learning at the Receiver Side}
\label{learn_receiver}
At the receiver side, we apply a GNN-based detector, i.e., $\rm{GNN}^{rx}$, to estimate the user message, consisting of EP and GNN modules, as shown in Fig. \ref{fig_e2e_train}. 
In the following, we explain the operations of these modules with the information of the learned constellation $\hat{\boldsymbol{\Omega}}$, received signal $\mathbf{y}$, channel matrix $\mathbf{H}$, and noise variance $\sigma^2$ in \eqref{eq_yhx_real}. In each EP iteration $t\in \{1,\cdots,T\}$, the GNN runs $L$ times.

\subsubsection{EP}
EP iteratively computes the cavity mean $\mathbf{x}_{{\rm obs}}^{(t)}$ variance $\mathbf{V}_{{\rm obs}}^{(t)}$ and posterior mean $\hat{\mathbf{x}}^{(t)}$ variance $\hat{\mathbf{V}}^{(t)}$ of the soft symbol estimates of $\mathbf{x}$. Firstly, the mean and covariance of the estimated symbols at iteration $t$ are calculated as 
\begin{subequations}\label{eq_obs1}
            \begin{equation} 
 \boldsymbol{\Sigma}^{(t)} = (\sigma^{-2}\mathbf{H}^{\top}\mathbf{H}+\boldsymbol{\lambda}^{(t-1)})^{-1}, 
            \end{equation}
            \begin{equation} 
\boldsymbol{\mu}^{(t)} = \boldsymbol{\Sigma}^{(t)}(\sigma^{-2}\mathbf{H}^{\top}\mathbf{y}+\boldsymbol{\gamma}^{(t-1)}),
            \end{equation} 
\end{subequations} 
where $\boldsymbol{\lambda}^{(t)}$ is a $2N_t\times 2N_t$ diagonal matrix with its diagonal value $\lambda^{(t)}_K>0$, $\boldsymbol{\gamma}^{(t)}=[\gamma^{(t)}_1,\dots,\gamma^{(t)}_K,\dots,\gamma^{(t)}_{2N_t}]$, and $K\in\{1,\dots,2N_t\}$. They are initialized as $\lambda^{(0)}_K=1, \gamma^{(0)}_K=0$. 
The mean and variance for symbol $x_K$ in $\mathbf{x}$ is obtained by using cavity probability, computed from \eqref{eq_obs1} \cite{Jespedes-TCOM14}, given as
\begin{subequations}\label{eq_obs2}  
            \begin{equation} \label{eq_obs2_1}  
 q^{(t)}(x_K) \propto \mathcal{N}(x_K:x_{{\rm obs},K}^{(t)}, v_{{\rm obs},K}^{(t)}),
            \end{equation}
where 
            \begin{equation} \label{eq_obs2_2}  
 v_{{\rm obs},K}^{(t)} = \frac{\Sigma_K^{(t)}}{{1-\Sigma_K}^{(t)}\lambda_K^{(t-1)}},  
            \end{equation}
            \begin{equation} \label{eq_obs2_3}  
x_{{\rm obs},K}^{(t)}  = v_{{\rm obs},K}^{(t)}(\frac{\mu_{K}^{(t)}}{\Sigma_K^{(t)}} - \gamma_{K}^{(t-1)}), 
            \end{equation} 
\end{subequations} 
where $\mu_{K}^{(t)}$ is the $K$-th element of $\boldsymbol{\mu}^{(t)}$, and $\Sigma_K^{(t)}$ is the $K$-th diagonal element of $\boldsymbol{\Sigma}^{(t)}$.
The posterior soft symbol estimate and its variance based on the cavity probability are given as
\begin{subequations}\label{eq_est}
    \begin{equation} 
    \hat{x}_K^{(t)}=   
    \begin{cases} 
    \displaystyle \sum_{\hat{a}_m\in\hat{\boldsymbol{\Omega}}} \Re(\hat{a}_m)   q^{(t)}(x_K=\Re(\hat{a}_m)),   & 1\!\leq\!K\!\leq\!N_t,\\ 
    \displaystyle \sum_{\hat{a}_m\in\hat{\boldsymbol{\Omega}}} \Im(\hat{a}_m)   q^{(t)}(x_K=\Im(\hat{a}_m)), & \rm{otherwise},
    \end{cases}  
    \end{equation}  

    \begin{equation} 
     \hat{v}_{K}^{(t)}=   
    \begin{cases} 
    \displaystyle \sum_{\hat{a}_m\in\hat{\boldsymbol{\Omega}}}\! (\Re(\hat{a}_m)\! -\!\hat{x}_{K}^{(t)})^2 q^{(t)}(x_K\!=\!\Re(\hat{a}_m)),     1\!\leq\!K\!\leq\!N_t,\\ 
   \displaystyle \sum_{\hat{a}_m\in\hat{\boldsymbol{\Omega}}}\! (\Im(\hat{a}_m)\!  -\!\hat{x}_{K}^{(t)})^2 q^{(t)}(x_K\!=\!\Im(\hat{a}_m)),  \rm{otherwise}.
    \end{cases}  
    \end{equation} 
\end{subequations}   
The soft symbol estimates are then concatenated as a vector as $\hat{\mathbf{x}}^{(t)}=[\hat{x}_1^{(t)},\dots,\hat{x}_K^{(t)},\dots,\hat{x}_{2N_t}^{(t)}]$. Its variance is represented by a diagonal matrix $\hat{\mathbf{V}}^{(t)}$ with diagonal values $[\hat{v}_1^{(t)},\dots,\hat{v}_K^{(t)},\dots,\hat{v}_{2N_t}^{(t)}]$.
Moment matching \cite{Jespedes-TCOM14} between the cavity mean $\mathbf{x}_{{\rm obs}}^{(t)}$ variance $\mathbf{V}_{{\rm obs}}^{(t)}$ and posterior mean $\hat{\mathbf{x}}^{(t)}$ variance $\hat{\mathbf{V}}^{(t)}$ is performed, resulting in an update of the precision matrix $\mathbf{\lambda}^{(t)}$ and mean $\mathbf{\gamma}^{(t)}$,
 \begin{subequations}\label{GEPNet_281}
            \begin{equation}\label{281a}
 \mathbf{\lambda}^{(t)}= (\hat{\mathbf{V}}^{(t)})^{-1}-(\mathbf{V}_{{\rm obs}}^{(t)})^{-1},
            \end{equation}
            \begin{equation}\label{281b}
\mathbf{\gamma}^{(t)} =(\hat{\mathbf{V}}^{(t)})^{-1} \hat{\mathbf{x}}^{(t)}-(\mathbf{V}_{{\rm obs}}^{(t)})^{-1}{\mathbf{x}}_{{\rm obs}}^{(t)},
            \end{equation} 
\end{subequations}    
where $\mathbf{x}_{{\rm obs}}^{(t)}\!=\![x_{{\rm obs},1}^{(t)},\dots,x_{{\rm obs},2N_t}^{(t)}]$, and diagonal matrix $\mathbf{V}_{{\rm obs}}^{(t)}$ with diagonal values $[v_{{\rm obs},1}^{(t)},\dots,v_{{\rm obs},2N_t}^{(t)}]$.
If $\mathbf{\lambda}^{(t)}$ yields a negative value, which should not be the case as it is an inverse variance term \cite{Alva_GEP},  we set $\mathbf{\lambda}^{(t)}=\mathbf{\lambda}^{(t-1)}$ and  $\mathbf{\gamma}^{(t)}=\mathbf{\gamma}^{(t-1)}$.
Finally, a damping operation is performed to smooth $\mathbf{\lambda}^{(t)}$ and $\mathbf{\gamma}^{(t)}$
\begin{subequations}\label{GEPNet_282}
            \begin{equation}\label{282a}
 \mathbf{\lambda}^{(t)}= (1-\eta)\mathbf{\lambda}^{(t)} + \eta \mathbf{\lambda}^{(t-1)},
            \end{equation}
            \begin{equation}\label{282b}
\mathbf{\gamma}^{(t)} =(1-\eta)\mathbf{\gamma}^{(t)} + \eta \mathbf{\gamma}^{(t-1)},
            \end{equation} 
\end{subequations}    
where $\eta $ is a weighting coefficient. $\mathbf{\lambda}^{(t)}$ and $\mathbf{\gamma}^{(t)}$ are then used by \eqref{eq_obs1} for the next iteration, and the whole process repeats for $T$ times.

\subsubsection{GNN}  
In each EP iteration, the GNN module is used to improve the cavity estimation, i.e., \eqref{eq_obs2_1} by taking $x_{{\rm obs},K}^{(t)}$ in \eqref{eq_obs2_2} and $ v_{{\rm obs},K}^{(t)}$ \eqref{eq_obs2_3} as inputs. In the GNN of $\rm{GNN}^{rx}$, the node $K$ represents the real-valued symbol $x_K$ in $\mathbf{x}$ and edge $\mathbf{e}_{JK}^{\rm{rx}}$ captures the interference relationship between nodes $J$ and $K$.
The GNN output is then forwarded to \eqref{eq_est} in EP as improved cavity probability.
The GNN module follows the GNN architecture presented in Section II, consisting of propagation, aggregation, and readout. 
The initial feature of node $K$ is expressed as 
\begin{equation}\label{eq_init_fea}
        \mathbf{u}_K^{(0),\rm{rx}} = \mathbf{W}_1 \cdot [\mathbf{y}^{\top} \mathbf{h}_K, \mathbf{h}_K^{\top} \mathbf{h}_K, \sigma^2]^{\top} + \mathbf{b}_1,
\end{equation}
where $\mathbf{W}_1 \in \mathbb{R}^{S_u^{\rm{rx}} \times 3}$ is the weight matrix, $\mathbf{b}_1 \in \mathbb{R}^{S_u^{\rm{rx}}}$ is the bias vector, and $S_u^{\rm{rx}}=8$ denotes the size of the node feature. $\mathbf{e}_{JK}^{\rm{rx}} =\left[ -\mathbf{h}_K^{\top} \mathbf{h}_J, \sigma^2 \right]^{\top}$ is the edge feature.
The node feature is then updated throughout the propagation \eqref{eq_prop} and aggregation \eqref{eq_agg}. In the aggregation module, \eqref{eq_agg} is executed with the prior information $\mathbf{r}_K = [x_{{\rm obs},K}^{(t)}, v_{{\rm obs},K}^{(t)}]^{\top}$ obtained from EP. After $L$ round message passing, the final node feature $\mathbf{u}_K^{(L),\rm{rx}} \in \mathbb{R}^{S_u^{\rm{rx}}}$ is fed to the readout to calculate the cavity probability of user messages. 
  
Here, we estimate the cavity probability of the user message, which is the same as the complex symbol probability. This differs from the cavity probability of the real and imaginary components of the symbol used in \cite{Alva_GEP}. The feature of message $s_k$ for user $k$ will need to be extracted from two nodes, i.e., $\mathbf{u}_k^{(L),\rm{rx}}$ and $\mathbf{u}_{k+N_t}^{(L),\rm{rx}}$.
Thus, we stack the feature of the first $N_t$ nodes with the other half, resulting in $\tilde{\mathbf{u}}_k^{(L),\rm{rx}} \in \mathbb{R}^{2S_u^{\rm rx}}= \left[ \left(\mathbf{u}_k^{(L),\rm{rx}}\right)^{\top}, \left(\mathbf{u}_{k+N_t}^{(L),\rm{rx}}\right)^{\top}\right]^{\top}$.  
The estimated cavity probability representing $k$-th user message at iteration $t$ is then obtained by using \eqref{eq_readout} in Section II, resulting in $\hat{\mathbf{s}}_k^{(t)}=[\hat{s}_{k,1}^{(t)},\dots,\hat{s}_{k,m}^{(t)},\dots,\hat{s}_{k,M}^{(t)}]$ that represents the normalized cavity probability of the estimated message for user $k$. This is then forwarded to \eqref{eq_est} in EP by setting $q^{(t)}(x_K\!=\!\Re(\hat{a}_m))=q^{(t)}(x_K\!=\!\Im(\hat{a}_m))=\hat{s}_{k,m}^{(t)}$, for $K=k$ or $K=k+N_t$.
After $T$ iterations of EP, the final output is given as $\hat{\mathbf{s}}_{1}^{(T)},\dots,\hat{\mathbf{s}}_{N_t}^{(T)}$.
 
The key difference between $\rm GNN^{rx}$ and \cite{Alva_GEP} lies in their readout processing: $\rm GNN^{rx}$ uses \eqref{eq_readout} to output the cavity probability of the message for each user from two nodes' features, whereas (26) in \cite{Alva_GEP} is used to readout from a single node feature. 

\subsection{E2E Loss}   
In offline training, we jointly optimize the NN parameters of the $\rm{GNN}^{tx}$ and $\rm{GNN}^{rx}$ by minimizing the CE loss of cavity probability estimation of the user messages via backpropagation, as shown in Fig. \ref{fig_e2e_train}, expressed as
\begin{equation}\label{eq_loss}
        Loss = -\frac{1}{BS}\sum_{bs=1}^{BS }\sum_{k=1}^{N_t }\sum_{m=1}^{M} {\mathbb{I}_{m = s_{k}}}  {\sf log}(\hat{s}_{k,m}^{(T)}),
\end{equation}
where $BS$ denotes the batch size, and $\mathbb{I}_{m = s_k}$ is an indicator function that takes value 1 if $m = s_k$ and 0 otherwise. 
Minimizing the loss \eqref{eq_loss} is equivalent to minimizing the SER for the MU-MIMO system. Here, the CE loss in \eqref{eq_loss} is based on the cavity probability of the message $\hat{s}_{k,m}^{(T)}$ in \eqref{eq_readout}. This is different from \cite{Alva_GEP}, which uses the cavity probability of the real and imaginary components of the symbol in the loss calculation.
During the offline training, all the NN parameters in $\rm{GNN}^{tx}$ and $\rm{GNN}^{rx}$ are optimized by using the Adam optimizer via backpropagation, as illustrated in Fig. \ref{fig_e2e_train}. 
 
\section{E2E architecture for online inference}  
\label{s_e2e_test} 
Once the offline training is completed, the optimized NN parameters in $\rm{GNN}^{rx}$ and the final learned constellation $\hat{\boldsymbol{\Omega}}$ obtained from $\rm{GNN}^{tx}$ are sent to the base station. The base station then broadcasts the final learned constellation to all users for online inference.
The E2E architecture used for online inference differs from that used for training, as user $k$ can map its message $s_k$ to modulated symbol $\tilde{x}_k$ via the final learned constellation $\hat{\boldsymbol{\Omega}}$ directly without executing the $\rm{GNN}^{tx}$. The complex symbol $\tilde{x}_k$ is then modeled as $\mathbf{x}$ in \eqref{eq_yhx_real}.

At the receiver, $\rm{GNN}^{rx}$ uses the final learned constellation $\hat{\boldsymbol{\Omega}}, \mathbf{y},\mathbf{H}$, and $\sigma^2$ to obtain the estimated message.
The final output of $\rm{GNN}^{rx}$ is the estimated normalized cavity probability of the user message, e.g., $\hat{\mathbf{s}}_k^{(T)}$ for user $k$. 
The index of $\hat{\mathbf{s}}_k^{(T)}$ with the maximum probability is then selected as the estimated message $\hat{s}_k$.

\section{Performance Evaluation}\label{S_sim_res} 
\subsection{Simulation Settings}\label{res_settings}
We consider two different MUI environments: $N_t=2, N_r=8$ for low MUI, and $N_t=N_r=8$ for high MUI. For a fair comparison, $M=16$ is used for both predefined and learned constellation points $\boldsymbol{\Omega}$ and $\hat{\boldsymbol{\Omega}}$, respectively. The predefined constellation is based on a standard $M$-ary quadrature amplitude modulation (16-QAM).
For the E2E configurations, we set $N_{p1}=128, N_{p2}=64, N_{u_1}=128, N_{r1}=N_{r2}=256, L=2, \eta=0.7$, and $T=10$.
The training of the proposed E2E is divided into $500$ epochs with $1000$ batches in each epoch, and each batch contains $128$ samples, i.e., $BS=128$. Each sample includes a realization of transmitted messages $s_1,\dots,s_{N_t}$, channel matrix $\mathbf{H}$, and noise $\mathbf{n}$. 
The NN parameters in E2E are jointly optimized using the Adam optimizer with a learning rate of $0.001$.  

The following schemes, based on a predefined 16-QAM modulator, are used for performance comparison: 
\textbf{1) MMSE:} A classical linear minimum-mean-square-error (MMSE) detector \cite{1327798}.  
\textbf{2) EP:} We use EP from \cite{Jespedes-TCOM14} with $T=10$ iterations.
\textbf{3) GEPNet:} We employ the state-of-the-art GNN-based detector GEPNet from \cite{Alva_GEP} with $T=10$ iterations, following the training procedure outlined therein.
\textbf{4) ML:} ML achieves optimal performance by exhaustively searching all possible combinations of transmitted symbols.

\subsection{Complexity Analysis} 
\label{S_complexity}
\begin{figure}
\centering
{\includegraphics[width=2.5in]{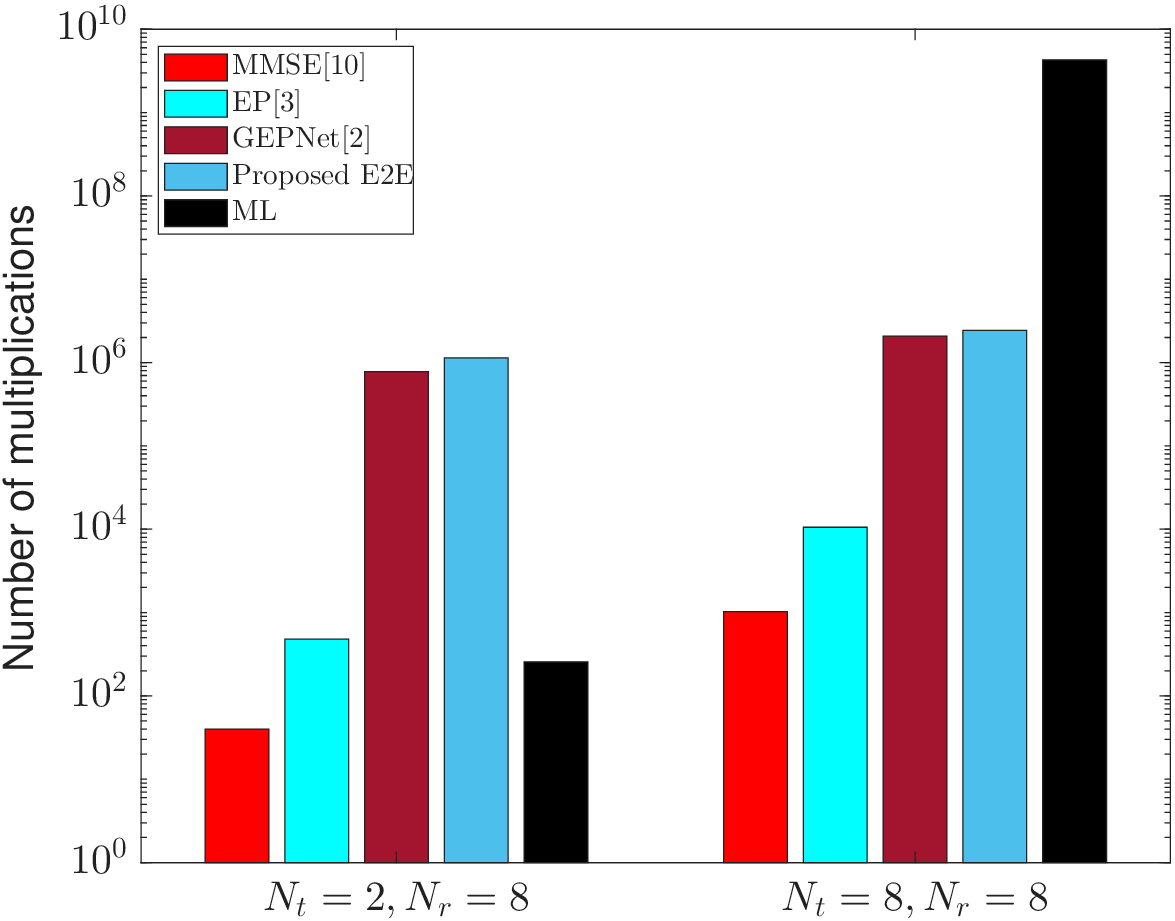}} 
\caption{Complexity comparison}
\vspace{-0.3cm}
\label{fig_complexity}
\vspace{-0.3cm}
\end{figure}

We use the order of the number of multiplications to evaluate the computational complexity of the proposed E2E during online inference for the real-valued system \eqref{eq_yhx_real}. The complexity comparison is summarized in Table \ref{Tab_complexity}.
The computational complexity of the proposed E2E lies at the receiver side only ($\rm GNN^{rx}$), as the $\rm GNN^{tx}$ is not executed during online inference, as discussed in Section \ref{s_e2e_test}. 
The differences between the $\rm GNN^{rx}$ and GEPNet are the readout \eqref{eq_readout} and posterior probability calculation \eqref{eq_est}, as discussed in Section \ref{s_e2e_train}. In $\rm GNN^{rx}$, the computational complexity of these modules is $\mathcal{O}(S_u^{\rm{rx}} N_{r1}+N_{r1}N_{r2}+N_{r2}M)N_tT$ and $\mathcal{O}(MN_tT)$, respectively. In contrast, in GEPNet, they are $\mathcal{O}(S_u^{\rm{rx}} N_{r1}+N_{r1}N_{r2}+N_{r2}\sqrt{M})N_tT$ and $\mathcal{O}(\sqrt{M}N_tT)$, respectively.
The differences in complexity between these modules are negligible compared to the overall computational complexity of GEPNet \cite{Alva_GEP}.     
Here, we provide numerical results for evaluating the complexity of the proposed scheme, as shown in Fig. \ref{fig_complexity}. As the number of users increases, the complexity of the proposed E2E is comparable to GEPNet and much lower than ML during online inference.

\begin{table}[!t]    
\setlength\tabcolsep{0.5pt}
\footnotesize
\caption{Complexity Comparison}
\centering
\resizebox{0.49\textwidth}{!}{
	\begin{tabular}{| c| c| c|}
		\hline  
        \textbf{Scheme}  & \textbf{Computational complexity}   	\\	\hline 
        MMSE\cite{1327798} &  $\mathcal{O}(N_t^3+N_t^2N_r)$  	\\ \hline
        EP \cite{Jespedes-TCOM14} 	&  $\mathcal{O}((N_t^3+N_t^2N_r+N_t\sqrt{M})T)$  	\\ \hline
        GEPNet \cite{Alva_GEP} 	& \makecell{$\mathcal{O}( (N_t^3+N_t^2N_r+N_t\sqrt{M}+S_u^{\rm{rx}}N_{r1}+N_{r1}N_{r2}+N_{r2}\sqrt{M}+ $\\$ (S_u^{\rm{rx}}N_{p1}+N_{p1}N_{p2}+N_{p2}S_u^{\rm{rx}}+N_{u1}S_u^{\rm{rx}})N_tL)T)$ } 	\\ \hline
        Proposed E2E    & 	 \makecell{$\mathcal{O}( (N_t^3+N_t^2N_r+N_tM+S_u^{\rm{rx}}N_{r1}+N_{r1}N_{r2}+N_{r2}M+$\\$(S_u^{\rm{rx}}N_{p1}+N_{p1}N_{p2}+N_{p2}S_u^{\rm{rx}}+N_{u1}S_u^{\rm{rx}})N_tL)T)$}	\\   \hline     
        ML	&   $\mathcal{O}(M^{N_t})$  \\  \hline
       
	\end{tabular} 
	\label{Tab_complexity}
    } 
\end{table}

\begin{figure}[!t] 
\centering
\subfloat[$N_t=2,N_r=8$ (Low MUI)]{\includegraphics[width=2.4in]{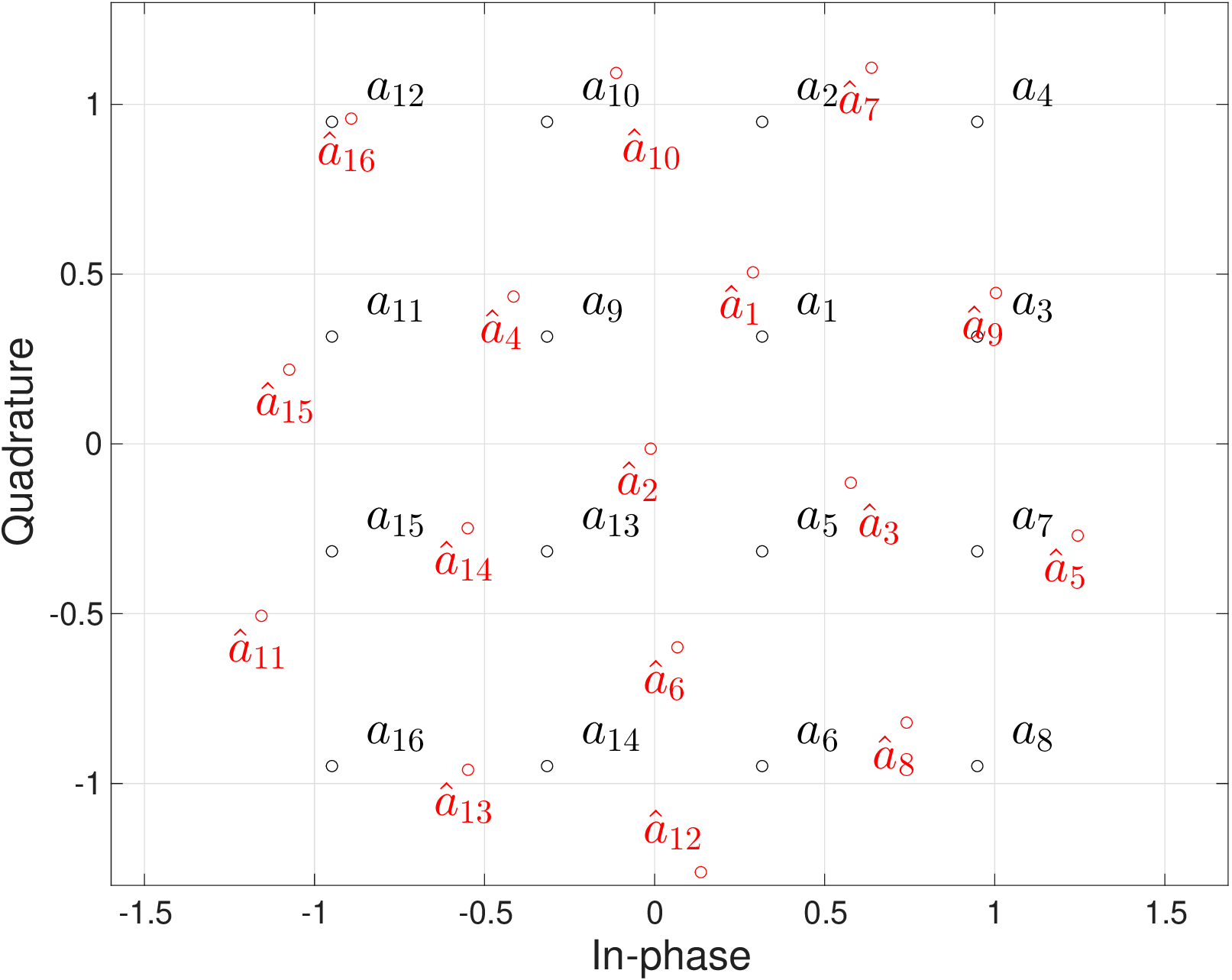}%
\label{fig_cons_28}} 
\vspace{-0.2cm}
\hfil 
\subfloat[$N_t=N_r=8$ (High MUI)]{\includegraphics[width=2.4in]{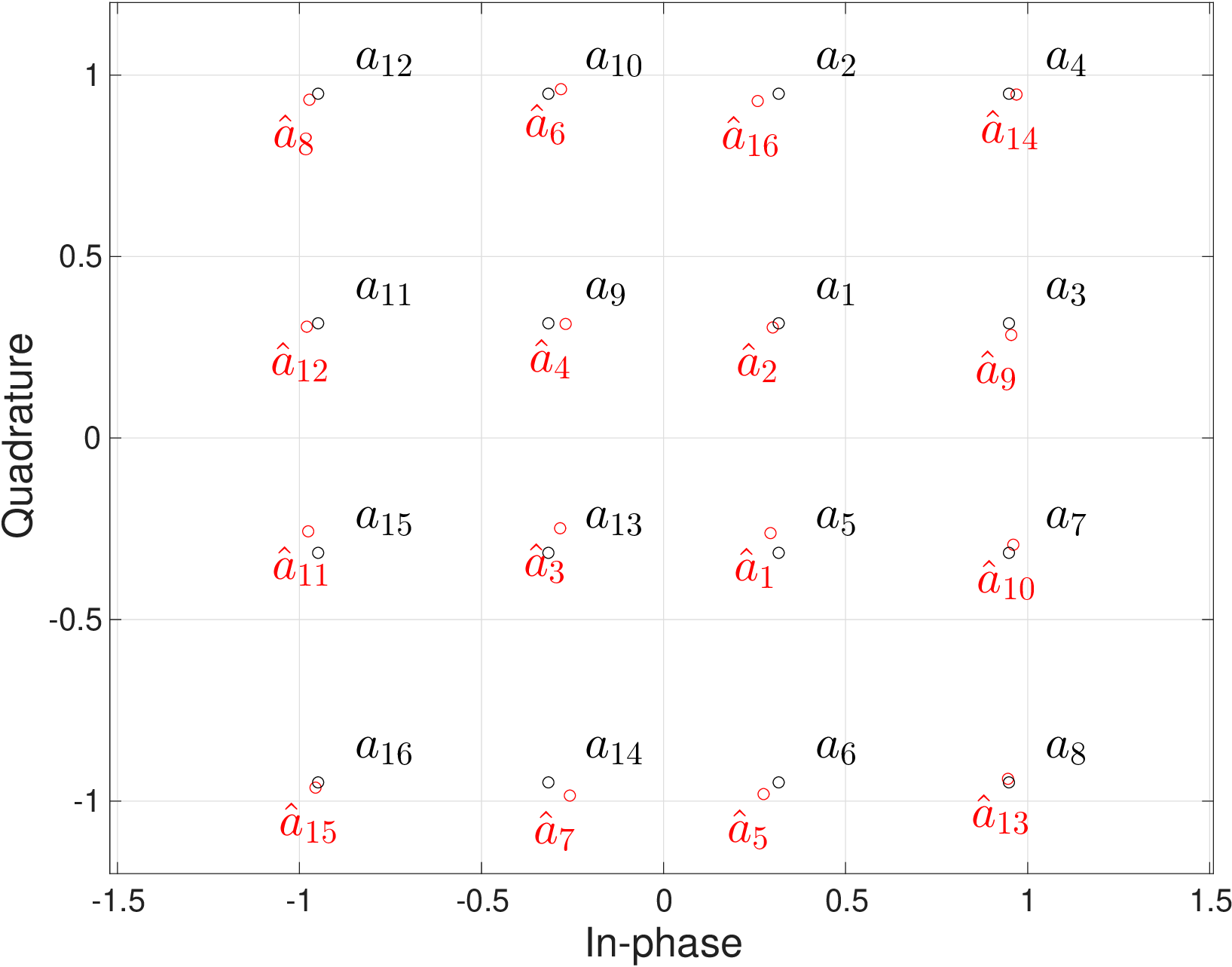}%
\label{fig_cons_44}}
\caption{Comparison of constellation $\hat{\boldsymbol{\Omega}}$ (${\color{red}\circ}$) and $\boldsymbol{\Omega}$  (${\color{black}\circ}$) in different MUI environments.} \label{fig_cons}
\vspace{-0.5cm}
\end{figure}

\subsection{Simulation Results}\label{res_res}
In this subsection, we first compare the final $M=16$ learned constellation points $\hat{\boldsymbol{\Omega}}$ generated by $\rm{GNN}^{tx}$ after training with 16-QAM constellation $\boldsymbol{\Omega}$ in different MUI environments.
Fig. \ref{fig_cons} illustrates the differences between the learned and predefined constellation points $\hat{a}_m, m\in\{1,\dots,M\}$ in $\hat{\boldsymbol{\Omega}}$ (in red color) and $a_m$ in $\boldsymbol{\Omega}$ (in black color).
In a low MUI environment, as illustrated in Fig. \ref{fig_cons_28}, the structure of the final learned constellation significantly differs from $16$-QAM. In contrast, in a high MUI environment, the final learned constellation retains the rectangular structure while it is still shifted, as shown in Fig. \ref{fig_cons_44}. For example, the symbol constellation points $a_1\in \boldsymbol{\Omega}$ and $\hat{a}_2\in \hat{\boldsymbol{\Omega}}$ that have similar phases and amplitudes are used to transmit different messages $s_k=1$ and $s_k=2$ for user $k$, respectively.
The adaptive behavior in different MUI environments demonstrates the ability of $\rm{GNN}^{tx}$ to adjust constellation patterns according to the MUI and receiver in the E2E system.

\begin{figure}[!t]
\centering
\subfloat[$N_t=2,N_r=8$ (Low MUI)] 
{\includegraphics[width=2.4in]{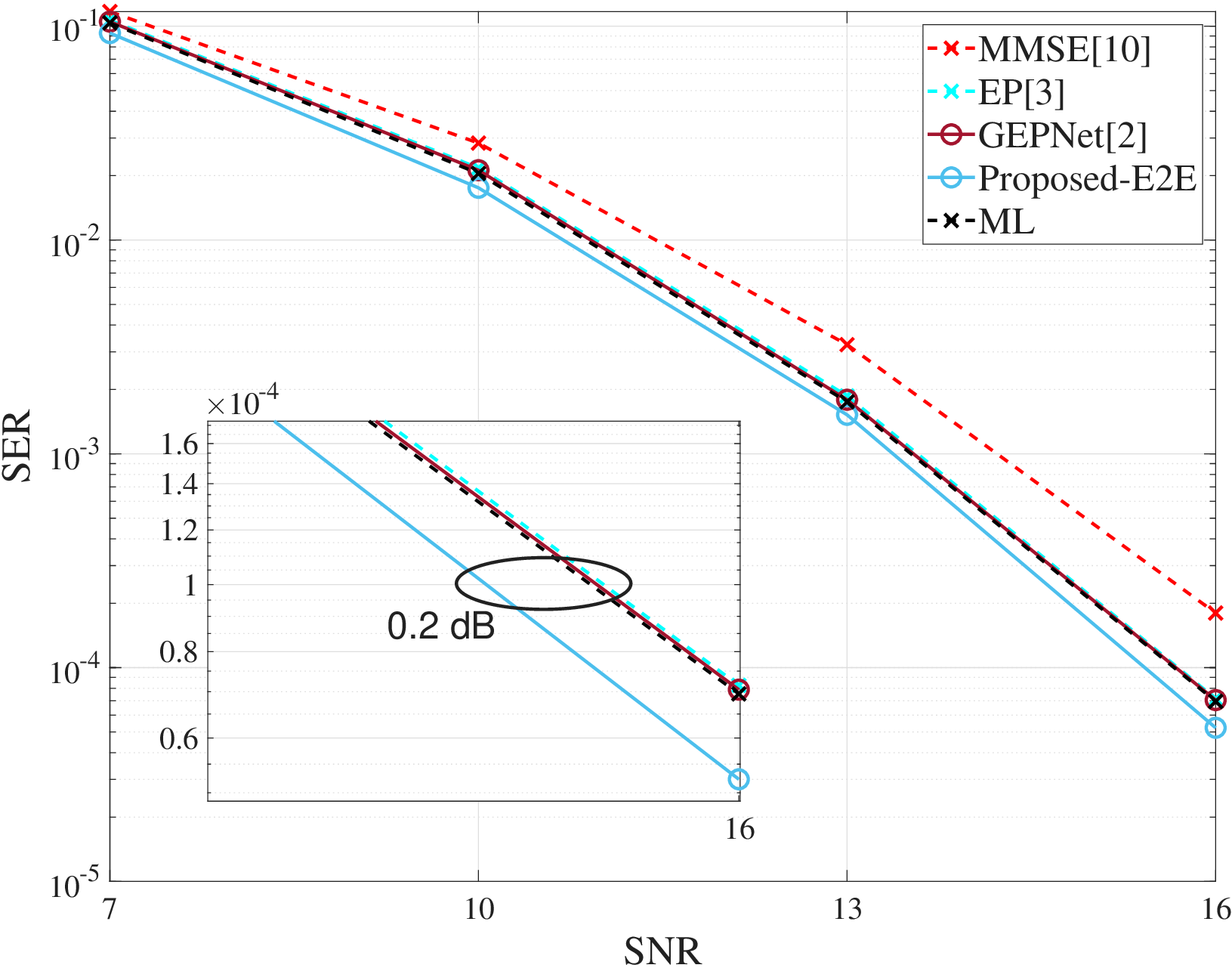}%
\label{fig_ser_28}} 
\vspace{-0.2cm}
\hfil
\subfloat[$N_t=N_r=8$ (High MUI)] 
{\includegraphics[width=2.4in]{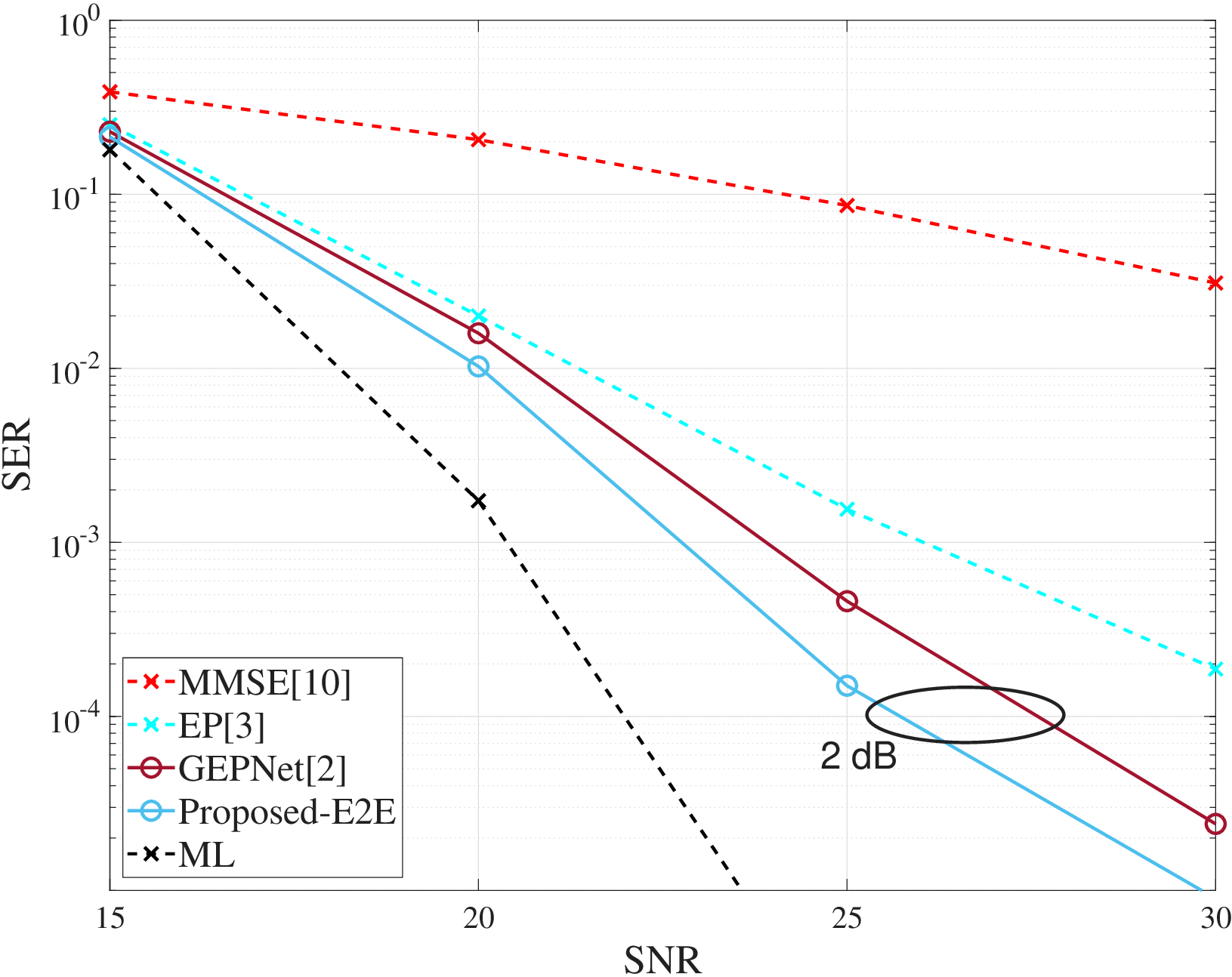}%
\label{fig_ser_88}} 
\caption{SER performance comparison in different MUI environments}
\vspace{-0.5cm}
\end{figure}

We then compare the SER performance of the proposed E2E with comparison schemes in different MUI environments. 
In a low MUI environment, Fig. \ref{fig_ser_28} shows that the proposed E2E outperforms all others, as joint optimization results in alignment between transmitter and receiver processes. The proposed E2E outperforms ML by around 0.2 dB at SER $=10^{-4}$, as the final learned constellation $\hat{\boldsymbol{\Omega}}$ used by E2E is better than $\boldsymbol{\Omega}$ used by ML. 
In a high MUI environment, Fig. \ref{fig_ser_88} demonstrates that the proposed E2E outperforms GEPNet by around 2 dB at SER $=10^{-4}$, and significantly outperforms EP and MMSE. Due to the high MUI, an SER gap still exists between the proposed E2E and ML.

\section{Conclusion}
\label{S_conclusion}

We proposed a GNN-based E2E learning approach for MU-MIMO that jointly optimizes a GNN-based modulator and a GNN-based detector. Simulation results showed that the SER performance of the proposed E2E outperforms GEPNet in different MUI environments and even surpasses ML under low MUI conditions, while maintaining a comparable complexity to GEPNet.


{\renewcommand{\baselinestretch}{1.1}
\begin{footnotesize}
\bibliographystyle{IEEEtran}
\bibliography{main}
\end{footnotesize}}

\end{document}